\documentclass[twocolumn,amsmath,amssymb,pre]{revtex4-1}

\usepackage{graphicx}
\usepackage{dcolumn}
\usepackage{bm}
 
\begin{document}

\title{Rigidity sensing explained by active matter theory}

\author{P.~Marcq}
\affiliation{
Physico-Chimie Curie, 
Institut Curie, CNRS, Universit\'e Pierre et Marie Curie,
26 rue d’Ulm, F-75248 Paris Cedex 05 France
}
\author{N.~Yoshinaga}
\affiliation{Fukui Institute for Fundamental Chemistry, Kyoto
University, Kyoto 606-8103, Japan}
\author{J.~Prost}
\affiliation{E.S.P.C.I., 10 rue Vauquelin, 75005 Paris, France}

\date{September 6, 2011}

\begin{abstract}
{
The magnitude of traction forces exerted by living animal cells 
on their environment is a monotonically increasing and approximately
sigmoidal function of the stiffness of the external medium.
This observation is rationalized using active matter theory:
adaptation to substrate rigidity results from an interplay 
between passive elasticity and active contractility.
}
\end{abstract}

\maketitle

Living cells respond to mechanical as well as biochemical cues.
Rigidity sensing designates the web of complex mechanisms
whereby a cell will adapt, as a function of 
the elastic modulus of its environment,
aspects of its phenotype  as diverse as its motility, 
gene expression,  proliferation,
or fate after differentiation \cite{Discher2005,Janmey2009}.

The traction forces that a cell exerts on a flat, elastic plate
depend in a non-trivial way on the extracellular stiffness
\cite{Saez2005}: roughly proportional 
to the elastic modulus in a softer environment, they saturate 
to a finite value for stiffer substrates \cite{Ghibaudo2008}.
Similar force-rigidity data are obtained whether forces are measured locally
\cite{Ghibaudo2008} or globally
\cite{Mitrossilis2009, Mitrossilis2010,Webster2011},
for integrin-mediated \cite{Saez2005,Ghibaudo2008} as well as
cadherin-mediated adhesion \cite{Ganz2006,Ladoux2010},
and even when the traction forces are exerted by 
assemblies of cells in a monolayer epithelium \cite{Saez2010}. 
Single-cell rheology assays \cite{Mitrossilis2010,Webster2011} 
show that the response of a cell to a sudden change of substrate rigidity is 
faster than the data acquisition rate allows to detect.
These observations call for a simple, generic explanation, 
valid on short time scales where cell signaling cannot operate.

In the context of adhesion-dependent mechanosensing, a
simple 'two-spring model' was introduced in \cite{Schwarz2006}, 
predicting that stiffer environments lead to stronger traction forces.
A 'three-spring model' was later proposed in order to explain 
the stiffness-dependent orientation of stress fibers 
in adherent cells \cite{Zemel2010}, where contractility 
modulates cytoskeletal stiffness via a
phenomenological polarizability coefficient.
In this Letter, we formulate and solve a simpler model  
derived from active matter theory, a generic description of living matter 
where the mechano-chemical transduction due to molecular motors (activity) 
plays a central role \cite{Kruse2005,Juelicher2007}.
We obtain a (static) force-rigidity relationship that agrees well with 
experimental data.
We give an expression of the (dynamic) loading rate, expected to be valid
on time scales too short for cytoskeletal remodeling and protein 
recruitment to occur ($t \ll 10^2$ s 
\cite{Icard-Arcizet2008,Allioux-Guerin2009}).

\section*{MODEL}

Active matter theory formulates constitutive equations
that take into account viscoelasticity, activity, the 
polar nature of cytoskeletal biopolymers \cite{Kruse2005},
and respect the principles of linear irreversible thermodynamics.
In one spatial dimension, we write the stress of an 
homogeneous, elastic and contractile material 
as the sum of an elastic and an active contribution:
$\sigma =   \sigma^{\mathrm{el}} + \sigma_A$.
The active stress $\sigma_A > 0$ 
is proportional to $\Delta \mu$, the difference of chemical potential 
between products and reagents of the chemical reaction responsible for 
mechanochemical transduction (ATP hydrolysis):
$\sigma_A = \zeta \ \Delta \mu$, where $\zeta$ is a material
parameter of the cytoskeleton \cite{Juelicher2007}.
For simplicity, we restrict our description of the
cytoskeleton to the \emph{linear} regime,
and include:
\begin{itemize}
\item[\emph{(i)}]{
\emph{elasticity}, described by a linear spring of 
length $l_C(t)$  at time $t$,
rest length $l^0_C$ and spring constant $k_C$;
} 
\item[\emph{(ii)}]{
\emph{activity}, modeled by an active force $F_A = \sigma_A \ S$
 across a section of area $S$.
} 
\end{itemize}
In the nonlinear regime, elastic moduli may also depend on activity,
as hypothesized in \cite{Zemel2010}.
Assuming, as in \cite{Kaunas2011}, that
the rest length of elastic cytoskeletal structures is time-dependent
due to motor activity is also beyond the scope of the linear regime
that we consider here.

\begin{figure}[t]
\vspace*{-10pt}
\centering{
\includegraphics[width=0.4\textwidth]{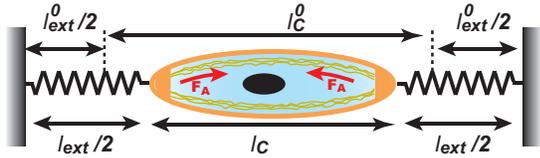}
}\vspace*{-10pt}
\caption{\label{fig:manip}
Schematic representation of the model}
\vspace*{-10pt}
\end{figure}

The extra-cellular environment is represented by a linear spring
of length $l_{\mathrm{ext}}(t)$  at time $t$,
rest length $l^0_{\mathrm{ext}}$, and spring constant $k_{\mathrm{ext}}$
(see Fig.~\ref{fig:manip}).
The traction force $F(t)$ exerted by the cell on its environment reads:
\begin{equation}
  \label{eq:force:spring}
F(t) =   k_{\mathrm{ext}} \left( l_{\mathrm{ext}}(t) - l^0_{\mathrm{ext}} \right).
\end{equation}
The sign is chosen as $F>0$ for contraction.
Under usual experimental conditions,
the total length $l_{\mathrm{tot}}$  of the system (cell + substrate)
is constant: $l_{\mathrm{tot}} = l_C(t) + l_{\mathrm{ext}}(t)$.
The force balance equation reads:
\begin{equation}
  \label{eq:forcebalance}
  -F(t) +  k_C \left( l_C(t) - l^0_C \right) + F_A = 0.
\end{equation}

\section*{STATICS}

The amplitude of the traction force at equilibrium reads:
\begin{equation}
  \label{eq:forcevsk}
  F^{\mathrm{eq}} = 
F_{\mathrm{Sat}} \; \frac{k_{\mathrm{ext}}}{k_{\mathrm{ext}} + k_C}.
\end{equation}
When $k_{\mathrm{ext}} \gg k_C$,
the traction force saturates to $F_{\mathrm{Sat}}$:
\begin{equation}
  \label{eq:finfty}
  F_{\mathrm{Sat}} =  F_A  + 
k_C \left( l_{\mathrm{tot}} - l_{\mathrm{ext}}^0 - l^0_C \right),
\end{equation}
the sum of $F_A$ and of residual stresses
$k_C ( l_{\mathrm{tot}} - l_{\mathrm{ext}}^0 - l^0_C )$.
We expect the ensemble average of residual stresses
to cancel: $< F_{\mathrm{Sat}} > = < F_A >$.  
As long as $k_{\mathrm{ext}} \ll k_C$, the traction force 
is a linear function of $k_{\mathrm{ext}}$:
$ F^{\mathrm{eq}} \simeq  (F_{\mathrm{Sat}}/k_C) \; k_{\mathrm{ext}}$.

\begin{figure}[t]
\vspace*{-15pt}
\centering{
\includegraphics[width=7cm,height=5cm]{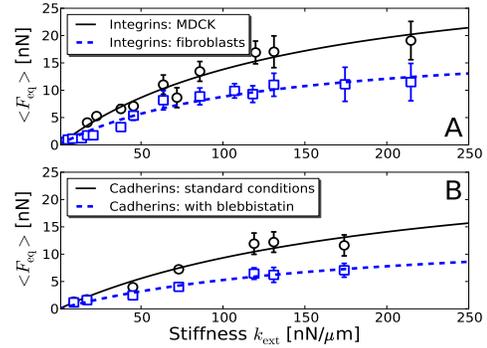}
}\vspace*{-10pt}
\caption{\label{fig:rig:fit}
We fit experimental data (average traction force per pillar, markers) 
with Eq.~(\ref{eq:forcevsk}) (lines).
\textbf{A}: integrin-mediated adhesions 
(Figs.~2a and 2c of \cite{Ghibaudo2008}).
MDCK cells (black circles): $F_{\mathrm{Sat}} = 34 \pm 4 \;\mathrm{nN}$, 
$k_C = 146 \pm 30 \; \mathrm{nN}/\mu\mathrm{m}$ (black solid line);
fibroblasts (blue squares): 
$F_{\mathrm{Sat}} = 19 \pm 3 \;  \mathrm{nN}$, 
$k_C = 119 \pm 35 \; \mathrm{nN}/\mu\mathrm{m}$ (blue dashed line).
\textbf{B}: cadherin-mediated adhesions (Fig.~4a of \cite{Ladoux2010}).
Standard conditions (black circles): 
$F_{\mathrm{Sat}} =  29 \pm 10 \;  \mathrm{nN}$, 
$k_C =  209 \pm 112 \; \mathrm{nN}/\mu\mathrm{m}$ (black solid line);
cells treated with blebbistatin (blue circles): 
$F_{\mathrm{Sat}} =  15 \pm 8 \;\mathrm{nN}$, 
$k_C =  183 \pm 164 \; \mathrm{nN}/\mu\mathrm{m}$ (blue dashed line).
Error bars correspond to $95 \%$ confidence levels. 
}
\vspace*{-5pt}
\end{figure}

Experimentally, a wide range of rigidities $k_{\mathrm{ext}}$
can be obtained when the substrate is a dense array of 
cylindrical elastomer micropillars whose stiffness depends 
on their radius and height.
Depending on the coating protein, traction forces are transmitted 
through integrin-mediated adhesions 
(with fibronectin) \cite{Saez2005,Ghibaudo2008}
or through cad\-he\-rin-mediated adhesions 
(with N-cad\-he\-rin) \cite{Ganz2006,Ladoux2010}.
Fig.~\ref{fig:rig:fit} show that,
in both cases, experimental data are well fitted
by the force-rigidity relation Eq.~(\ref{eq:forcevsk}).
Note that the fitted values of saturation forces $F_{\mathrm{Sat}}$ 
were not observed: they correspond to  values of $k_{\mathrm{ext}}$
so large that the deflections of pillars would fall below
the experimental spatial resolution.

The data encompasses three cell types, 
Madin-Darby Canine Kidney (MDCK) cells, 3T3 fibroblasts,
and C2 mouse myogenic cells. 
Cytoskeletal organisation and adhesive properties of the cells vary 
substantially according to the type of adhesions and the
range of substrate rigidity. 
More diffuse cortical actin dominates when the environment is softer,
wheras actomyosin bundles are preferentially formed at high\-er 
rigidities \cite{Ghibaudo2008,Ladoux2010}.
In all cases, our simple model captures the essence of the 
force-rigidity dependence, and sums up biological variation 
into two quantitative parameters, the asymptotic traction
force $F_{\mathrm{Sat}}$ and the cytoskeletal stiffness $k_C$.
The order of magnitude of the saturation force 
$F_{\mathrm{Sat}} \sim 10 \; \mathrm{nN}$ corresponds
to an  active stress of the order of $10^4$ Pa.
We obtain this value upon
neglecting possible residual stresses, and using
$\sigma_A \sim  F_{\mathrm{Sat}}  / S$,
where the section $S$ of a micropillar is of order
$S \sim 1 \; \mu\mathrm{m}^2$. 
The cytosketal rigidity is of the order of
$10^2 \; \mathrm{nN}/\mu\mathrm{m}$,
corresponding to elastic moduli of the order of $10^5$ Pa,
a value intermediate between moduli typical of cortical 
actin \cite{Wottawah2005} and of stress fibers \cite{Deguchi2006}
(we used $E \sim k_C \; d/S$ with $d \sim 1 \; \mu\mathrm{m}$).

When cells are treated with blebbistatin, an inhibitor of contractility,
the value of $F_{\mathrm{Sat}}$ is  halved (Fig.~\ref{fig:rig:fit} B). 
Inspection of Eq.~(\ref{eq:finfty}) suggests that traction forces remain
non-zero due to residual activity of myosin motors, as proposed in
\cite{Ladoux2010}, or to non-zero residual stresses, or to a combination
of both effects. We note that the value of $k_c$ is almost unchanged.
However, assuming as in \cite{Zemel2010}
that activity modulates cytoskeletal stiffness via a
polarizability coefficient $\alpha$ leads to a ratio
$F^{\mathrm{eq}}/F_{\mathrm{Sat}} = 
k_{\mathrm{ext}}/ \left( k_{\mathrm{ext}} + (1 + \alpha) \; k_C \right)$
that depends through $\alpha$ on cytoskeletal contractility.
Experimental data \cite{Ladoux2010} suggests that
$F^{\mathrm{eq}}/F_{\mathrm{Sat}}$ is independent 
of the level of contractility,
in agreement with our prediction, Eq.~(\ref{eq:forcevsk}).

\section*{DYNAMICS}

The loading rate exerted by T cells immediately after 
receptor engagement with a model antigen-presenting cell 
was measured thanks to a biomembrane force set-up where micropipette aspiration 
controls the external rigidity, and found  to be linear in $k_{\mathrm{ext}}$
\cite{Husson2011}. 
Motivated by this result, we turn to the dynamics of traction forces, and
modify Eq.~(\ref{eq:forcebalance}) by taking into account
internal protein friction in a \emph{linear} force-velocity relationship:
\begin{equation}
  \label{eq:fv}
    F_A(t) =  F_S + \xi \; \frac{\mathrm{d}l_C}{\mathrm{d}t}
\end{equation}
where $F_S$ is the stall force, and $\xi$ is a friction coefficient
\cite{Tawada1991}. 
Eliminating other variables in Eq.~(\ref{eq:forcebalance}), 
we obtain a differential equation for the traction force:
\begin{equation}
  \label{eq:dfdt}
  \frac{\mathrm{d}F}{\mathrm{d}t} + \frac{F}{\tau} = 
  \frac{F^{\mathrm{eq}}}{\tau} 
\end{equation}
with a viscoelastic time 
$\tau = \xi/\left(k_{\mathrm{ext}} + k_C \right)$.
Integration from an initial time $t_0$ with initial force
$F(t = t_0) = F^0$ gives
$  F(t) = F^{\mathrm{eq}} +
\left(F^0  - F^{\mathrm{eq}} \right) \; e^{-(t-t_0)/\tau}$.
For zero initial force, we find that the
initial loading rate is proportional to the
substrate rigidity:
\begin{equation}
  \label{eq:dfdt_t0}
  \frac{\mathrm{d}F}{\mathrm{d}t}_{|(t - t_0) \ll \tau} \simeq 
 \frac{F_{\mathrm{Sat}}}{\xi} \; k_{\mathrm{ext}},
\end{equation}
in agreement with  \cite{Husson2011}, 
where the initial time $t_0$ is set when pulling starts so that $F^0 = 0$.
We checked that Eq.~(\ref{eq:dfdt_t0}) still holds if we replace
the linear force-velocity equation (\ref{eq:fv}) by Hill's 
equation \cite{Hill1938}.
Since $F_{\mathrm{Sat}}$ is a function of activity, 
we predict that the loading-rate-rigidity data
will be modified upon 
treatment with contractility agonists and antagonists.

\section*{CONCLUSION AND OUTLOOK}

At low external rigidity, cell traction forces increase linearly
with the stiffness of the substrate. Their 
constant ratio was first interpreted as a displacement regulated  by the
cell \cite{Saez2005}. We show that regulation is not necessary to explain the
force-rigidity relationship.
Within the framework of linear irreversible thermodynamics, 
we propose a minimal model 
who\-se consequences are consistent with available experimental data. 
We predict that both the saturation force $F_{\mathrm{Sat}}$,
exerted for large stiffness,
and the constant displacement $F_{\mathrm{Sat}}/k_C$, observed 
at low stiffness, depend upon the contractility level.
Our description is relevant for several types of adhesive structures
and cytoskeletal organisation. 
Unlike \cite{Schwarz2006}, we ignore the dynamics of adhesive contacts
through which force is transmitted to the substrate.
Other monotonically increasing functions of stiffness that depend
on two parameters also fit the same
experimental data: it is our hope that this work will foster 
further quantitative experiments to confirm -- or disprove --
our model.

To treat the dynamics of traction forces, we include internal 
friction, and obtain an initial loading rate proportional
to external stiffness, as observed experimentally \cite{Husson2011}. 
In accord with single-cell rheology assays 
\cite{Mitrossilis2010,Webster2011},
the loading rate $\mathrm{d}F/\mathrm{d}t$ responds instantaneously
to variations of $k_{\mathrm{ext}}$.

The merit of our analysis is to show that the simplest equations of 
active matter dictated by symmetry and conservation laws are sufficient 
to describe a behavior which might seem to require a more elaborate 
regulation at first sight. This  suggests that other features such 
as stress fiber diameter and equilibrium with the rest of the 
actin-myosin system could be described within the general framework 
of active gels.
We hope that extensions of our model will allow to understand 
quantitatively how more complex cell processes depend on extracellular rigidity.
Including membrane elasticity and cortical tension in an appropriate 
geometry may explain why the initial loading rate exerted by T cells 
upon receptor engagement 
saturates for stiffer environements \cite{Husson2011}.
The dynamics of wetting of the microplate by the cell 
must be taken into account
to describe traction forces exerted during cell spreading
\cite{Mitrossilis2009, Mitrossilis2010}.
Finally, biochemical signaling, protein recruitment and 
remodeling of adhesive and cytoskeletal structures act over
longer time scales  \cite{Icard-Arcizet2008,Allioux-Guerin2009}
and may enhance the mechanical effects described here.

\section*{ACKNOWLEGMENTS}

The authors thank A.~Asnacios, F.~Graner, J.-F.~Joanny, J.~Husson,
B.~La\-doux and P.~Silberzan for fruitful discussions.
This work was supported by JSPS, MAEE and MESR under the Japan-France 
Integrated Action Program (SAKURA).

\end{document}